\documentclass[10pt,aps,twocolumn,nofootinbib,superscriptaddress,eqsecnum]{revtex4-2}
\usepackage[normalem]{ulem}
\usepackage{cancel}
\usepackage[T1]{fontenc}
\usepackage{amsmath,amssymb,bm,amsthm}
\usepackage{bbm}
\usepackage{graphicx,xcolor}
\usepackage{tikz}
\usepackage[colorlinks=true, allcolors=blue]{hyperref}
\usepackage{orcidlink}
\newtheorem{proposition}{Proposition}
\newcommand{\Tr}{\mathrm{Tr}}
\newcommand{\id}{\bm{I}}
\newcommand{\ket}[1]{\left|#1\right\rangle}
\newcommand{\bra}[1]{\left\langle#1\right|}
\newcommand{\proj}[1]{\ket{#1}\!\bra{#1}}

\begin{document}
\title{Finite-size Entanglement Certification via Third-Order Local Randomized Measurements}

\author{Giovanni Scala~\orcidlink{0000-0003-2685-0946}}
\affiliation{Institute of Physics, Faculty of Physics, Astronomy and Informatics,
Nicolaus Copernicus University, Grudziądzka 5/7, 87-100 Toru\'{n}, Poland}
\affiliation{Dipartimento Interateneo di Fisica, Politecnico di Bari, 70126 Bari, Italy}
\email{giovanni.scala@poliba.it}
\affiliation{INFN, Sezione di Bari, 70126 Bari, Italy}

\author{Corrado C. Mattiacci~\orcidlink{0009-0005-4554-5260}}
\email{corrmatt@doktorant.umk.pl}
\affiliation{Institute of Physics, Faculty of Physics, Astronomy and Informatics,
Nicolaus Copernicus University, Grudziądzka 5/7, 87-100 Toru\'{n}, Poland}
\author{Dorota Pietryka~\orcidlink{0009-0004-1947-9426}}
\affiliation{Faculty of Mathematics, Physics and Informatics, University of Gda\'{n}sk,Wita Stwosza 57, 80-308 Gda\'{n}sk,
Poland}

\author{Gniewomir Sarbicki~\orcidlink{0000-0002-6403-8839}}
\affiliation{Institute of Physics, Faculty of Physics, Astronomy and Informatics,
Nicolaus Copernicus University, Grudziądzka 5/7, 87-100 Toru\'{n}, Poland}
\affiliation{Institute of Advanced Studies, \\
Nicolaus Copernicus University in Toruń, Wileńska 4, 87-100 Toruń, Poland}

\begin{abstract}
\noindent Certifying entanglement in high-dimensional systems usually requires full state tomography, whose cost grows rapidly with the system dimension. \emph{Local} randomized measurements offer a scalable alternative, but existing tests based on second-order correlations access only limited information about the state. Here, we derive a finite-size entanglement certificate that extends local randomized measurements to third order. The additional third-order information reveals entanglement that remains undetected at second order, while a dimension-independent concentration bound provides rigorous control of finite-sample errors. Our result opens a practical route to extracting stronger entanglement information from experimental platforms without the dimension-dependent overhead of state tomography.
\end{abstract}
\maketitle
\section{Introduction}
\noindent Entanglement is a defining feature of quantum theory and a resource for quantum computation, communication, and sensing~\cite{Leone2025,HHHH,GuhneToth2009,chruscinski2014entanglement}. Any protocol that relies on this resource must guarantee that entanglement is present under realistic noise and with a finite amount of data. Such validation often requires dedicated measurement rounds that could otherwise be used for the target task~\cite{cieslinski2024analysing,Ghoreishi2025,Elben2022,LogicCHSH2026,miller2026detecting}. Entanglement certification should therefore provide reliable conclusions while adding as little experimental overhead as possible \cite{Friis2019}.

Full reconstruction of the quantum state is not required for this purpose. Ideally, entanglement should be certified through a directly measurable quantity whose violation provides an unambiguous conclusion~\cite{terhal2002detecting,Huang2020,Aaronson2020,Nguyen2022,Guehne2007}. Established separability criteria, including positivity under partial transposition (PPT)~\cite{woronowicz1976positive,peres1996separability,Horodecki1996}, are powerful when the density operator is known. Applying them experimentally, however, generally requires reconstructing the state~\cite{MauroDAriano2003,cramer2010efficient,Christandl2012}. For a system of effective dimension $d_{\rm eff}$, this means estimating $O(d_{\rm eff}^3)$ parameters. Even optimized tomography therefore becomes increasingly demanding as the Hilbert-space dimension grows~\cite{Haah2017,Kanter2023,ChenLiLiu2024,LoweNayak2025,ChoKim2025,Bavaresco2018}.

\emph{Local} randomized measurements offer a non-tomographic alternative. Random local unitaries are followed by measurements in a fixed basis, allowing nonlinear properties of the state to be estimated directly from single-copy data~\cite{Elben2018,Vermersch2018,Rico2024,vanEnk2012PRL,guhne2024,guhne2023exp,Lami2026}. Existing entanglement tests of this type rely mainly on second-order quantities such as global and marginal purities\cite{brydges2019probing}. These measurements are experimentally accessible, but they capture only part of the correlations contained in the state and may leave entanglement undetected~\cite{brydges2019probing,Zhang2024,Vermersch2024}.

This raises a central question: \emph{can higher-order correlations provide stronger entanglement certification without restoring the dimension-dependent cost of tomography?} We answer this question by projecting the reduction criterion onto a fixed operator subspace and expressing the resulting certificate through local randomized measurements up to third order. The construction applies to finite experimental data and comes with a concentration bound that has no explicit dependence on the local Hilbert-space dimensions. It therefore allows experimental platforms to access entanglement information beyond second order while preserving the scalability that makes randomized measurements attractive.
\section{Projective Reduction Criterion}
We consider a bipartite state
$\rho\in\mathcal{B}(\mathbb{C}^{d_A}\otimes\mathbb{C}^{d_B})$
and denote its reduced states by $\rho_A=\Tr_B(\rho)$ and
$\rho_B=\Tr_A(\rho)$. Our starting point is the reduction map applied on one or the other subsystem~\cite{Horodecki1999,Cerf1999}
\begin{equation*}
\mathfrak{R}_A(\rho):=\rho_A\otimes\id_B-\rho,
\qquad
\mathfrak{R}_B(\rho):=\id_A\otimes\rho_B-\rho .
\end{equation*}
If $\rho$ is separable (it can be written as
$\rho=\sum_k p_k\,\sigma_k^A\otimes\sigma_k^B$, where $p_k\geq0$, $\sum_kp_k=1$, and $\sigma_k^A$ and $\sigma_k^B$
are local density operators) implies
\begin{align}
\mathfrak{R}_A(\rho)
&=
\sum_k p_k\,\sigma_k^A\otimes(\id_B-\sigma_k^B)
\succeq0,
\nonumber\\
\mathfrak{R}_B(\rho)
&=
\sum_k p_k\,(\id_A-\sigma_k^A)\otimes\sigma_k^B
\succeq0
\label{eq:RA_RB_separable_positive}
\end{align}
which means that:
\begin{equation*}
    \forall X \succeq0 \qquad \Tr\left(X\mathfrak{R}_{A/B}(\rho)\right) \succeq0.
\end{equation*}
Therefore, violation of either condition certifies entanglement. Because the
reduction criterion is weaker than PPT, it does not detect every entangled
state; however, any state that violates it is also distillable~\cite{Horodecki1999,jivulescu2015thresholds,Horodecki1998}.
Directly testing the positivity of $\mathfrak{R}_{A/B}(\rho)$ requires access
to the full operator space, and we will show that it would not be suitable for scalable
certification. 
We instead test its positivity within the fixed operator subspace generated by the operator tuple $V=\{\id,\rho_A\otimes\id_B,\id_A\otimes\rho_B,\rho\}$, thus taking $X$ to be the squared combinations of $V$.
Then for every real vector $\mathbf{a}\in\mathbb{R}^4$,
\begin{align}
0&\preceq
\Tr\!\left[
\left(\sum_{i=0}^3 a_iV_i\right)^2
\mathfrak{R}_{A/B}(\rho)
\right]
:=
\mathbf{a}^{ T}\mathfrak{M}_{A/B}(\rho)\mathbf{a} ,
\label{eq:Malpha_def}\\
&\mathfrak{M}_{A/B,ij}(\rho)
=
\Tr\!\left[
\frac{\{V_i,V_j\}}{2}\,
\mathfrak{R}_{A/B}(\rho)
\right],
\quad
i,j=0,\dots,3.
\nonumber
\end{align}
Thus, separability implies 
$$\mathfrak{M}_A(\rho)\succeq0, \qquad
\mathfrak{M}_B(\rho)\succeq0.$$
Consequently, a negative eigenvalue of either matrix certifies entanglement.
This construction replaces a positivity test on the full Hilbert space by projecting the reduction criteria into $V$ \cite{Yi2026}. 
Their entries are polynomials of degree at most three in $\rho$ and its
marginals, which makes them accessible through local randomized measurements up to
third order as we will show below. 
\section{Experimental Reconstruction of the Moment Matrices}
Having reduced the full reduction criterion to two fixed-size moment matrices $\mathfrak{M}_{A/B}$, we now show how their entries can be reconstructed from \emph{local} randomized measurements.
We first describe the measurement protocol and the accessible second- and third-order invariants. We then address the only term that is not directly measurable through a local protocol.
\subsection{Randomized measurements}
\label{sec:Randomized_measurements}
The entries of \(\mathfrak M_{A/B}(\rho)\) depend only on second- and third-order invariants, which can be reconstructed from local randomized measurements.
Ref.~\cite{QST2025} shows how those functions can be estimated via local randomized measurements.
Concretely, two parties $A,B$ share a bipartite quantum system in a state $\rho$.
At a realized local setting $u=u^A\otimes u^B$, each party measures $u\rho u^\dagger$ projectively in its local computational basis $\{\ket{i^{(A/B)}}\}_{i=1}^{d_{A/B}}$.
Throughout, the state $\rho$ is fixed.  For $s=1,\ldots,N_U$, let
\begin{align}
U_s&=U_s^A\otimes U_s^B,
\quad
U_1,\ldots,U_{N_U}
\overset{\mathrm{i.i.d.}}{\sim}
\mu_{\mathrm H}^A\otimes\mu_{\mathrm H}^B,
\nonumber\\
\mathbf U&=(U_1,\ldots,U_{N_U}),
\label{eq:random_unitary_model}
\end{align}
where uppercase $U_s$ denotes a random setting and lowercase $u_s$ one of its realizations. Here and below the Haar measure is $\mu_{\mathrm H}^A\otimes\mu_{\mathrm H}^B$. Set $\Omega=[d_A]\times[d_B]$.  For every setting $s$ and shot $r=1,\ldots,N_S$, let $I_{s,r}\in\Omega$ be the random measurement outcome.  Conditioned on $U_s=u$,
\begin{align}
\mathbb P_\rho(I_{s,r}=i\mid U_s=u)=p_\rho(i\mid u),
\nonumber\\
I_{s,1},\ldots,I_{s,N_S}\mid(U_s=u)
\overset{\mathrm{i.i.d.}}{\sim}p_\rho(\,\cdot\mid u).
\label{eq:conditional_outcome_model}
\end{align}
Conditioned on $\mathbf U$, the outcome samples belonging to different settings are independent.  We write $\mathbb P$ for probability, $\mathbb E[X]$ for expectation over all experimental randomness, and $\mathbb E_U[f(U)]$ for Haar expectation over the unitary only.  The notation $\mathbb E[X \mid U]$ denotes a random variable depending on $U$, whereas $\mathbb E[X\mid U=u]$ is its value at a fixed realization.  The subscripts in $\mathbb P_\rho$ and $\mathbb E_\rho$ only record the fixed state $\rho$ where the random variable is $I_{s,r}$; and $\mathbb E_U$, $\operatorname{Cov}_U$ when the random variable refers to the Haar-distributed unitary alone.
For $i=(i^{(A)},i^{(B)})\in\Omega$ and a fixed realization $u=u^A\otimes u^B$, the conditional probability is
\begin{equation*}
    p_\rho(i\mid u) = \Tr\!\left[ \left( \bigotimes_{\ell \in \{A,B\}} u^{(\ell)\dagger} \ket{i^{(\ell)}}\bra{i^{(\ell)}} u^{(\ell)} \right) \rho \right] .
\end{equation*}
For fixed outcomes $i_1,i_2,i_3\in\Omega$, the Haar-averaged third-order correlator is
\begin{equation}
    \bar{p}_\rho(i_1,i_2,i_3)
    =\mathbb E_U\!\left[
    p_\rho(i_1\mid U)p_\rho(i_2\mid U)p_\rho(i_3\mid U)
    \right] ,
    \label{eq:3order_corr}
\end{equation}
By Weingarten calculus~\cite{Collins2006,Puchala2017}, the set of all possible outcome triples partitions into exactly 10 discrete equivalence classes (orbits) $\mathcal{C}_0, \dots, \mathcal{C}_9$, determined by local index-equality patterns. In each subsystem, three indices $i_1, i_2, i_3$ can be all distinct ($[123]$), two equal ($[21]$), or all equal ($[1^3]$). 
Pairing these 3 local patterns across subsystems $A$ and $B$ yields 9 combinations; the $([21],[21])$ combination further splits into parallel $(\parallel)$ and cross $(\times)$ configurations based on whether the matching pair indices align across parties. Explicitly,
\begin{align*}
\mathcal C_0 &= ([1^3],[1^3]), & \mathcal C_1 &= ([1^3],[21]), & \mathcal C_2 &= ([1^3],[123]), \nonumber\\
\mathcal C_3 &= ([21],[1^3]), & \mathcal C_4 &= ([21],[21])_{\parallel}, & \mathcal C_5 &= ([21],[21])_{\times}, \nonumber\\
\mathcal C_6 &= ([21],[123]), & \mathcal C_7 &= ([123],[1^3]), & \mathcal C_8 &= ([123],[21]), \nonumber\\
\mathcal C_9 &= ([123],[123]).\label{eq:orbit_classe}
\end{align*}
For later use, define the deterministic indicators
\begin{equation*}
h_\mu(i_1,i_2,i_3)
=\mathbbm 1[(i_1,i_2,i_3)\in\mathcal C_\mu],
\qquad \mu=0,\ldots,9.
\end{equation*}
\begin{figure}[htbp]
\centering
\begin{tikzpicture}[scale=.33]
	\fill (0,0) circle (4pt); \fill (1,0) circle (4pt); \fill (2,0) circle (4pt);
	\fill (0,-1) circle (4pt); \fill (1,-1) circle (4pt); \fill (2,-1) circle (4pt);
	\node[below] at (1,-1.2) {$\mathcal C_0$};
\end{tikzpicture}
\hspace{.4cm}
\begin{tikzpicture}[scale=.33]
	\fill (0,0) circle (4pt); \fill (1,0) circle (4pt); \fill (2,0) circle (4pt);
	\fill (0,-1) circle (4pt); \fill (1,-1) circle (4pt); \fill (2,-1) circle (4pt);
	\draw[line width=1pt] (0,0) -- (1,0);
	\node[below] at (1,-1.2) {$\mathcal C_3$};
\end{tikzpicture}
\hspace{.4cm}
\begin{tikzpicture}[scale=.33]
	\fill (0,0) circle (4pt); \fill (1,0) circle (4pt); \fill (2,0) circle (4pt);
	\fill (0,-1) circle (4pt); \fill (1,-1) circle (4pt); \fill (2,-1) circle (4pt);
	\draw[line width=1pt] (0,0) -- (1,0) -- (2,0);
	\node[below] at (1,-1.2) {$\mathcal C_7$};
\end{tikzpicture}
\hspace{.4cm}
\begin{tikzpicture}[scale=.33]
	\fill (0,0) circle (4pt); \fill (1,0) circle (4pt); \fill (2,0) circle (4pt);
	\fill (0,-1) circle (4pt); \fill (1,-1) circle (4pt); \fill (2,-1) circle (4pt);
	\draw[line width=1pt] (0,-1) -- (1,-1);
	\node[below] at (1,-1.2) {$\mathcal C_1$};
\end{tikzpicture}
\hspace{.4cm}
\begin{tikzpicture}[scale=.33]
	\fill (0,0) circle (4pt); \fill (1,0) circle (4pt); \fill (2,0) circle (4pt);
	\fill (0,-1) circle (4pt); \fill (1,-1) circle (4pt); \fill (2,-1) circle (4pt);
	\draw[line width=1pt] (0,-1) -- (1,-1) -- (2,-1);
	\node[below] at (1,-1.2) {$\mathcal C_2$};
\end{tikzpicture}
\par\vspace{0.4cm}
\begin{tikzpicture}[scale=.33]
	\fill (0,0) circle (4pt); \fill (1,0) circle (4pt); \fill (2,0) circle (4pt);
	\fill (0,-1) circle (4pt); \fill (1,-1) circle (4pt); \fill (2,-1) circle (4pt);
	\draw[line width=1pt] (0,0) -- (1,0); \draw[line width=1pt] (0,-1) -- (1,-1);
	\node[below] at (1,-1.2) {$\mathcal C_4$};
\end{tikzpicture}
\hspace{.4cm}
\begin{tikzpicture}[scale=.33]
	\fill (0,0) circle (4pt); \fill (1,0) circle (4pt); \fill (2,0) circle (4pt);
	\fill (0,-1) circle (4pt); \fill (1,-1) circle (4pt); \fill (2,-1) circle (4pt);
	\draw[line width=1pt] (0,0) -- (1,0); \draw[line width=1pt] (1,-1) -- (2,-1);
	\node[below] at (1,-1.2) {$\mathcal C_5$};
\end{tikzpicture}
\hspace{.4cm}
\begin{tikzpicture}[scale=.33]
	\fill (0,0) circle (4pt); \fill (1,0) circle (4pt); \fill (2,0) circle (4pt);
	\fill (0,-1) circle (4pt); \fill (1,-1) circle (4pt); \fill (2,-1) circle (4pt);
	\draw[line width=1pt] (0,0) -- (1,0) -- (2,0); \draw[line width=1pt] (0,-1) -- (1,-1);
	\node[below] at (1,-1.2) {$\mathcal C_8$};
\end{tikzpicture}
\hspace{.4cm}
\begin{tikzpicture}[scale=.33]
	\fill (0,0) circle (4pt); \fill (1,0) circle (4pt); \fill (2,0) circle (4pt);
	\fill (0,-1) circle (4pt); \fill (1,-1) circle (4pt); \fill (2,-1) circle (4pt);
	\draw[line width=1pt] (0,0) -- (1,0); \draw[line width=1pt] (0,-1) -- (1,-1) -- (2,-1);
	\node[below] at (1,-1.2) {$\mathcal C_6$};
\end{tikzpicture}
\hspace{.4cm}
\begin{tikzpicture}[scale=.33]
	\fill (0,0) circle (4pt); \fill (1,0) circle (4pt); \fill (2,0) circle (4pt);
	\fill (0,-1) circle (4pt); \fill (1,-1) circle (4pt); \fill (2,-1) circle (4pt);
	\draw[line width=1pt] (0,0) -- (1,0) -- (2,0); \draw[line width=1pt] (0,-1) -- (1,-1) -- (2,-1);
	\node[below] at (1,-1.2) {$\mathcal C_9$};
\end{tikzpicture}
\caption{Diagrammatic representation of the 10 index-equality classes $\mathcal{C}_0, \dots, \mathcal{C}_9$; top and bottom rows represent subsystem $A$ and $B$ measurement outcomes across three copies. Unlinked nodes represent distinct indices; linked nodes represent equal indices.}
\label{fig:equality_classes}
\end{figure}
Let $I_1,I_2,I_3$ denote three generic outcomes that are conditionally i.i.d. under a common setting.  For a fixed unitary realization $u$, define the conditional class probabilities
\begin{align}
Y_\mu(u)
=&\mathbb E_\rho\!\left[h_\mu(I_1,I_2,I_3)\mid U=u\right],
\nonumber\\
\mathbf Y(u)=&\bigl(Y_0(u),\ldots,Y_9(u)\bigr)^{\mathsf T},
\label{eq:conditional_class_probabilities}
\end{align}
Here $\rho$ is fixed and is not a source of randomness.  More explicitly, $\mathbb E_\rho[\;\cdot\mid U=u]$ averages only over the three conditionally independent measurement outcomes with common conditional law $p_\rho(\,\cdot\mid u)$, so that
\begin{align}
\mathbb E_\rho\!&\left[h_\mu(I_1,I_2,I_3)\mid U=u\right]
=\nonumber\\
&\sum_{(i_1,i_2,i_3)\in\Omega^3}
h_\mu(i_1,i_2,i_3)
\prod_{r=1}^{3}p_\rho(i_r\mid u).
\label{eq:Ehmu}
\end{align}
Their deterministic Haar average $\mathbf Y=\mathbb E_U[\mathbf Y(U)]$. $\mathbb E_U$ averages only over the random local-unitary setting $U=U^A\otimes U^B$ distributed according to $\mu_{\mathrm H}^A\otimes\mu_{\mathrm H}^B$; hence
\begin{equation}
    \mathbf Y
=
\mathbb E_U[\mathbf Y(U)]
=
\int \mathbf Y(u)\,
\mathrm d\!\left(\mu_{\mathrm H}^A\otimes\mu_{\mathrm H}^B\right)(u).
\label{eq:EUY}
\end{equation}
Thus $\mathbf Y(u)$ is deterministic for fixed $u$, whereas $\mathbf Y(U)$ is random solely through the Haar-distributed setting $U$. By local-unitary invariance, $\bar p_\rho(i_1,i_2,i_3)$ has a common value $y_\mu$ for all $(i_1,i_2,i_3)\in\mathcal C_\mu$.  With $\mathbf y=(y_0,\ldots,y_9)^{\mathsf T}$ and with the class order fixed in Fig.~\eqref{fig:equality_classes}, the relation between representative correlators and class probabilities is
\begin{equation}
\mathbf Y=K\mathbf y,
\qquad
K=\operatorname{diag}(k_0,\ldots,k_9).
\label{eq:Y_K_y}
\end{equation}
Here $k_\mu=|\mathcal C_\mu|$ is the cardinality of the corresponding class:
\begin{align*}
k_0 & = d_A(d_A-1)(d_A-2) d_B(d_B-1)(d_B-2), \\ 
k_1 & = 3 d_A(d_A-1) d_B(d_B-1)(d_B-2), \\
k_2 & = d_A(d_A-1)(d_B-2) d_B, \\
k_3 & = 3d_A(d_A-1) d_B(d_B-1)(d_B-2), \\
k_4 & = 3d_A(d_A-1)d_B(d_B-1) \\
k_5 & = 6d_A(d_A-1)d_B(d_B-1) 
\\
k_6 & = 3d_A(d_A-1)d_B, \quad
k_7  = d_Ad_B(d_B-1)(d_B-2), \\
k_8 & = 3d_Ad_B(d_B-1), \quad
k_9  = d_Ad_B
\end{align*}
For example, if we take the 10th equality $\mathcal{C}_9$ there are only $k_9=d_A d_B$ possibilities such that $i_1^{(k)}= i_2^{(k)}=i_3^{(k)}$ for $k=A,B$.
By applying standard Weingarten identities~\cite{Collins2006,elben2019statistical,Knips2020}, one obtains a reconstruction map $L\in\mathbb R^{9\times10}$ such that
\begin{equation}
    \mathbf{x}=L\mathbf y=LK^{-1}\mathbf Y,
    \qquad L\in\mathbb R^{9\times10},
    \label{eq:inversion}
\end{equation}
where the vector $\mathbf x=(x_1,\ldots,x_8,x_S)^{\mathsf T}$ of local unitary invariants has entries~\cite{QST2025}

\begin{align}
x_1&=\Tr(\rho_B^2),\, x_2=\Tr(\rho_B^3),\quad x_3=\Tr(\rho_A^2),\quad x_5=\Tr(\rho^2),\nonumber\\
x_4&=\Tr\!\big[(\rho_A\otimes\rho_B)\rho\big],\quad
x_6=\Tr\!\big[\rho_B\,\Tr_A(\rho^2)\big],
x_7=\Tr(\rho_A^3),\quad \nonumber \\
x_8&=\Tr\!\big[\rho_A\,\Tr_B(\rho^2)\big],\quad x_S
= x_9+x_{10}, 
\label{eq:invariants}
\end{align}
where $x_9=\Tr(\rho^3)$ and $x_{10}=\Tr[(\rho^{T_A})^3]$.
Remarkably, this protocol is tomography-free and basis independent. It assumes trusted local random unitaries drawn from Haar measure or a suitable $t$-design, followed by fixed-basis readout~\cite{webb2016,Zhu2017,Ketterer2020}.
\subsection{Reconstruction of Moment matrices}
\label{sec:moment_matrices}
Evaluating Equation (\ref{eq:Malpha_def}) one can express the entries of $\mathfrak M_A$ by the invariants (\ref{eq:invariants}):
\begin{align}
\mathfrak{M}_{A,00}&=\Tr(\id\,\mathfrak{R}(\rho))=\Tr(\rho_A\otimes\id_B)-\Tr\rho=d_B-1,
\nonumber\\
\mathfrak{M}_{A,01}&=\Tr\big((\rho_A\otimes\id_B)\mathfrak{R}(\rho)\big)
=\Tr(\rho_A^2\otimes\id_B-(\rho_A\otimes\id_B)\rho\big)\nonumber\\
&=(d_B-1)\Tr\rho_A^2=(d_B-1)x_3,
\nonumber\\
\mathfrak{M}_{A,02}&=\Tr\big((\id_A\otimes\rho_B)\mathfrak{R}\big)
=\Tr(\rho_A\otimes\rho_B-(\id_A\otimes\rho_B)\rho\big)\nonumber\\
&=1-\Tr\rho_B^2=1-x_1,\nonumber\\
\mathfrak{M}_{A,03}&=\Tr(\rho\,\mathfrak{R})=\Tr\big(\rho(\rho_A\otimes\id_B)\big)-\Tr(\rho^2)\nonumber\\
&=\Tr\rho_A^2-\Tr\rho^2=x_3-x_5.
\nonumber\\
\mathfrak{M}_{A,11}
&=\Tr\big((\rho_A^2\otimes\id_B)\mathfrak{R}\big)
=\Tr(\rho_A^3\otimes\id_B-(\rho_A^2\otimes\id_B)\rho\big)\nonumber\\&
=(d_B-1)\Tr\rho_A^3=(d_B-1)x_7,
\nonumber
\end{align}
\begin{align}
\mathfrak{M}_{A,22}
&=\Tr\big((\id_A\otimes\rho_B^2)\mathfrak{R}\big)
=\Tr(\rho_A\otimes\rho_B^2-(\id_A\otimes\rho_B^2)\rho\big)\nonumber\\&
=\Tr\rho_B^2-\Tr\rho_B^3=x_1-x_2,\nonumber\\
\mathfrak{M}_{A,12}
&=\Tr\Big(\frac{\{\rho_A\otimes\id_B,\id_A\otimes\rho_B\}}{2}\mathfrak{R}\Big)
=\Tr\big((\rho_A\otimes\rho_B)\mathfrak{R}\big)\nonumber\\&
=\Tr(\rho_A^2\otimes\rho_B)-\Tr\big((\rho_A\otimes\rho_B)\rho\big)=x_3-x_4,\nonumber
\\
\mathfrak{M}_{A,13}
&=\Tr\Big(\frac{\{\rho_A\otimes\id_B,\rho\}}{2}\mathfrak{R}\Big)
=\Tr\big(\rho_A^3\big)-\Tr\!\big[\rho_A\,\Tr_B(\rho^2)\big]\nonumber\\&=x_7-x_8,
\nonumber\\
\mathfrak{M}_{A,23}
&=\Tr\Big(\frac{\{\id_A\otimes\rho_B,\rho\}}{2}\mathfrak{R}\Big)\nonumber\\&
=\Tr\!\big[(\rho_A\otimes\rho_B)\rho\big]-\Tr\!\big[\rho_B\,\Tr_A(\rho^2)\big]=x_4-x_6.
\nonumber
\end{align}
These identities follow from expanding the anticommutators and repeatedly using cyclicity of the trace together with the basic
partial-trace rules
$\Tr[(X_A\otimes \id_B)\rho]=\Tr(X_A\rho_A)$ and $\Tr[(\id_A\otimes Y_B)\rho]=\Tr(Y_B\rho_B)$.
Finally,
\begin{align}
\mathfrak{M}_{A,33}
&=\Tr\Big(\frac{\{\rho,\rho\}}{2}\mathfrak{R}(\rho)\Big)
=\Tr\big(\rho^2(\rho_A\otimes\id_B)\big)-\Tr(\rho^3)\nonumber\\
&
=\Tr\!\big[\rho_A\,\Tr_B(\rho^2)\big]-\Tr(\rho^3)=x_8-x_9.\label{eq:sm_M33}
\end{align}
Similarly for the entries of $\mathfrak{M}_B$.

The only obstruction to full measurability is the global cubic moment $x_9=\Tr(\rho^3)$ in Eq.~\eqref{eq:sm_M33}, which is not directly accessible in a purely local third-order randomized-measurement protocol~\cite{QST2025}. To retain a purely local protocol, we exploit that if $\rho$ is separable, then $\rho^{T_A}$ is also separable and positive. Hence, if $\rho$ is separable,
\begin{equation}
\bar{\mathfrak{M}}_{A/B}(\rho)
:=
\frac12\left[
\mathfrak{M}_{A/B}(\rho)+
\mathfrak{M}_{A/B}(\rho^{T_A})
\right]\succeq 0.
\label{eq:Mbar_alpha_def}
\end{equation}
This is a crucial observation because it leads to the following matrices, where each entry can be measured by the aforementioned randomised measurement protocol,
\begin{align}
\bar{\mathfrak{M}}_A(\rho)=
\begin{pmatrix}
d_B-1 & (d_B-1)x_3 & 1-x_1 & x_3-x_5 \\
\cdot & (d_B-1)x_7 & x_3-x_4 & x_7-x_8 \\
\cdot & \cdot & x_1-x_2 & x_4-x_6 \\
\cdot & \cdot & \cdot & x_8- \frac 12 x_S
\end{pmatrix},
\nonumber
\\
\bar{\mathfrak{M}}_B(\rho)=
\begin{pmatrix}
d_A-1 & (d_A-1)x_1 & 1-x_3 & x_1-x_5 \\
\cdot & (d_A-1)x_2 & x_1-x_4 & x_2-x_6 \\
\cdot & \cdot & x_3-x_7 & x_4-x_8 \\
\cdot & \cdot & \cdot & x_6- \frac 12 x_S
\end{pmatrix}
\label{eq:sm_MbarExplicit_correct}
\end{align}
where $\cdot$ denotes symmetric completion. Therefore the minimal eigenvalue of $\bar{\mathfrak{M}}_A(\rho) \oplus \bar{\mathfrak{M}}_B(\rho)$, namely
\begin{equation}
\mathcal{E}(\rho)
:=
\min\!\left\{
\lambda_{\min}\!\big(\bar{\mathfrak{M}}_A(\rho)\big),\,
\lambda_{\min}\!\big(\bar{\mathfrak{M}}_B(\rho)\big)
\right\}
\label{eq:Esym_def}
\end{equation}
when it is negative, then it signals that $\rho$ is entangled.

Although this criterion is weaker than the reduction criterion in Eq.~\eqref{eq:RA_RB_separable_positive}, the finite-sample bounds below have no explicit local-dimension dependence at fixed normalized tolerance and confidence.  Sign-certification cost also depends on the true normalized margin, which may depend on dimension.
\section{Finite-sample entanglement certification}
To formulate the criterion for finite experimental data with Chebyshev bound, we first  express the moment matrices $\bar{\mathfrak{M}}_{A/B}$ as linear functions of the measured class-probability vector $\mathbf Y$ and then renormalize both the signal $\mathcal{E}$ and its statistical fluctuations to correctly derive the measurement overhead \(N_{\rm tot}\) required to certify entanglement at fixed accuracy and confidence.
\subsection{Linear reconstruction and normalization}
Given $\bar{\mathfrak{M}}_{A/B}$ a real symmmetric $4\times 4$ matrix, we can define $\mathbf m_{A/B}:=\mathrm{svec}(\bar{\mathfrak{M}}_{A/B})\in \mathbb R^{10}$, where
\begin{multline}
    \operatorname{svec}(\bar{\mathfrak{M}})=(\bar{\mathfrak{M}}_{00},\sqrt{2}\bar{\mathfrak{M}}_{01},\sqrt{2}\bar{\mathfrak{M}}_{02}, \sqrt{2}\bar{\mathfrak{M}}_{03}, \\ \bar{\mathfrak{M}}_{11},\sqrt{2}\bar{\mathfrak{M}}_{12},\sqrt{2}\bar{\mathfrak{M}}_{13},\bar{\mathfrak{M}}_{22},\sqrt{2}\bar{\mathfrak{M}}_{23},\bar{\mathfrak{M}}_{33})^{T}
\end{multline}
so that $\|\mathrm{svec}(X)\|_2:=\sqrt{\sum_i\mathrm{svec}(X)_i^2}=\sqrt{\mathrm{Tr}X^TX}=:\|X\|_{\rm HS}$, where $\|\cdot \|_{HS}$ denotes the Hilbert-Schmidt norm.
Therefore Eq.~\eqref{eq:sm_MbarExplicit_correct} can be written linearly as follows,
\begin{equation}
 \mathbf m_{A/B}=A_{\bar{\mathfrak{M}}_{A/B}} \mathbf x+  \mathbf b_{A/B},
 \label{eq:mvec}
\end{equation}
where the multiplicative coefficients are collected in $A_{\bar{\mathfrak{M}}_{A/B}} \in \mathbb{R}^{10 \times 9}$ and $\mathbf b_{A/B}$ explicitly written in App. \ref{app:explicit_linear_maps}.
Using Eq. \eqref{eq:inversion} we get
\begin{equation}
\mathbf m_\alpha=B_{\alpha,d}\mathbf Y+\mathbf b_\alpha,
\quad
B_{\alpha,d}:=A_{\bar{\mathfrak M}_\alpha}LK^{-1}
\in\mathbb R^{10\times10},
\label{eq:main_B_map}
\end{equation}
The deterministic reconstruction maps $B_{\alpha,d}$ can amplify fluctuations in the reconstructed class-probability vector.  To obtain one Lipschitz constant valid for both orientations, define
\begin{equation}
C_\alpha(d_A,d_B):=
\|B_{\alpha,d}\|_{\rm op}.
\label{eq:C_normalization}
\end{equation}
With this normalization,
\begin{equation}
\frac{\|B_{\alpha,d}v\|_2}{C_{\alpha}(d_A,d_B)}\leq\|v\|_2,
\qquad \alpha\in\{A,B\},\quad v\in\mathbb R^{10}.
\label{eq:B_normalization_contraction}
\end{equation}
The rescaled quantities are
\begin{equation}
\widetilde{\bar{\mathfrak{M}}}_{A/B}
:=
\frac{\bar{\mathfrak{M}}_{A/B}}{C_{A/B}(d_A,d_B)},
\qquad
\widetilde{\mathcal{E}}_{A/B}(\rho)
:=
\frac{\mathcal{E}_{A/B}(\rho)}{C_{A/B}(d_A,d_B)}.
\label{eq:normalized_Esym}
\end{equation}
Because 
$C_{A/B}(d_A,d_B)$
is deterministic, the normalization preserves the sign
\begin{align*}
\bar{\mathfrak{M}}_{A/B}\succeq0 
\iff \widetilde{\bar{\mathfrak{M}}}_{A/B}\succeq0,
\quad{\mathcal{E}}(\rho)<0
 \iff \widetilde{\mathcal{E}}(\rho)<0,
\end{align*}
where $\widetilde{\mathcal{E}}(\rho)=\min\{\widetilde{\mathcal{E}}_{A}(\rho),\widetilde{\mathcal{E}}_{B}(\rho)\}$. This rescaling removes a \emph{deterministic} dimension-dependent amplification factor, therefore the deviation threshold $\epsilon$, which must take into account only the statistical errors can be compared directly to normalized fluctuations.

We can now determine the number of measurements required to certify entanglement from finite data. To this end, we bound the deviation of the estimated normalized witness $\widehat{\widetilde{\mathcal{E}}}$ from its true value $\widetilde{\mathcal{E}}$. If the resulting confidence interval lies strictly below zero, the state is certified as entangled at the corresponding confidence level. We begin with a conservative estimate based on Chebyshev's inequality.

\subsection{Chebyshev bound}
Let for the setting $s$: $\mathbf Z_{s,abc}
:=\bigl(Z_{s,abc}^{(0)},\ldots,Z_{s,abc}^{(9)}\bigr)^{\mathsf T}$ where
\begin{equation}
Z_{s,abc}^{(\mu)}
:=h_\mu(I_{s,a},I_{s,b},I_{s,c}), \text{ for }
1\leq a<b<c\leq N_S.
\label{eq:class_variables}
\end{equation}
Each $\mathbf Z_{s,abc}$ takes values in $\{\mathbf e_0,\ldots,\mathbf e_9\}$.  The all-triples estimator at setting $s$ is
\begin{align}
&\widehat{\mathbf Y}_{N_S}^{(s)}
:=\binom{N_S}{3}^{-1}
\sum_{1\leq a<b<c\leq N_S}\mathbf Z_{s,abc},
\nonumber\\
&\mathbb E_\rho\!\left[
\widehat{\mathbf Y}_{N_S}^{(s)}\mid U_s
\right]=\mathbf Y(U_s).
\label{eq:app_general_estim}
\end{align}
The conditional expectation is taken only over the measurement outcomes $I_{s,a}$, $I_{s,b}$, and $I_{s,c}$ at setting \(s\), while \(U_s\) is held fixed.  More explicitly, for every realization \(U_s=u\) and every \(\mu=0,\ldots,9\), using Eq.\eqref{eq:Ehmu},
\begin{align}
\mathbb E_\rho&\!\left[
\bigl[\widehat{\mathbf Y}_{N_S}^{(s)}\bigr]_\mu
\mid U_s=u
\right]
=\\
&
\binom{N_S}{3}^{-1}
\sum_{1\leq a<b<c\leq N_S}
\mathbb E_\rho\!\left[
h_\mu(I_{s,a},I_{s,b},I_{s,c})
\mid U_s=u
\right]\nonumber\\
&=
\sum_{(i_1,i_2,i_3)\in\Omega^3}
h_\mu(i_1,i_2,i_3)
\prod_{r=1}^{3}p_\rho(i_r\mid u)
=Y_\mu(u).\nonumber
\end{align}
Thus \(\widehat{\mathbf Y}_{N_S}^{(s)}\) remains random at fixed \(u\) because it depends on the sampled outcomes, whereas its conditional mean \(\mathbf Y(u)\) is deterministic.  Accordingly,
\(\mathbb E_\rho[\widehat{\mathbf Y}_{N_S}^{(s)}\mid U_s]
=\mathbf Y(U_s)\)
is an equality of random vectors depending only on \(U_s\).

The global estimator and its deterministic target are
\begin{align}
&\widehat{\mathbf Y}_{N_U,N_S}
:=\frac1{N_U}\sum_{s=1}^{N_U}\widehat{\mathbf Y}_{N_S}^{(s)},
\nonumber\\
&\mathbf Y:=\mathbb E_U[\mathbf Y(U)],
\quad
\mathbb E[\widehat{\mathbf Y}_{N_U,N_S}]=\mathbf Y.
\label{eq:app_global_estimator_general}
\end{align}
Here \(\mathbb E_U\) averages only over a single random setting \(U=U^A\otimes U^B\) with law \(\mu_{\mathrm H}^A\otimes\mu_{\mathrm H}^B\) as in Eq. \eqref{eq:EUY}.
By contrast, the unconditional expectation
\(\mathbb E\,[\widehat{\mathbf Y}_{N_U,N_S}]\) averages over the complete hierarchical experiment at fixed \(\rho\): first the i.i.d.\ settings \(U_1,\ldots,U_{N_U}\) are sampled from the local Haar measure, and then the measurement outcomes are sampled according to their conditional distributions \(p_\rho(\,\cdot\mid U_s)\). Indeed, by the chain property of expectation value and the conditional unbiasedness in Eq.~\eqref{eq:app_general_estim},
\begin{align}
\mathbb E\,&\!\left[
\widehat{\mathbf Y}_{N_U,N_S}
\right]
=
\mathbb E_U\!\left[
\mathbb E_\rho\!\left[
\widehat{\mathbf Y}_{N_U,N_S}\mid\mathbf U
\right]
\right]\nonumber\\
&=
\mathbb E_U\!\left[
\frac1{N_U}\sum_{s=1}^{N_U}\mathbf Y(U_s)
\right]
=
\frac1{N_U}\sum_{s=1}^{N_U}
\mathbb E_U[\mathbf Y(U_s)]
=\mathbf Y.\nonumber
\end{align}
Thus \(\widehat{\mathbf Y}_{N_U,N_S}\) is random through both the sampled settings and the measurement outcomes, whereas \(\mathbf Y\) is a deterministic vector determined by the fixed state \(\rho\) and the Haar measurement ensemble.

The normalized moment-matrix estimator is defined by
\begin{equation}
\operatorname{svec}\!\left(
\widehat{\widetilde{\bar{\mathfrak M}}}_\alpha
\right)
:=\frac{B_{\alpha,d}\widehat{\mathbf Y}_{N_U,N_S}+\mathbf b_\alpha}
{C_\alpha(d_A,d_B)},
\qquad \alpha\in\{A,B\}.
\label{eq:estimated_normalized_matrix}
\end{equation}
Thus $\widehat{\mathbf Y}_{N_U,N_S}$ and the moment matrices reconstructed linearly from it are unbiased.  By contrast,
\begin{equation}
\widehat{\widetilde{\mathcal E}}
:=\min_{\alpha\in\{A,B\}}
\lambda_{\min}\!\left(
\widehat{\widetilde{\bar{\mathfrak M}}}_\alpha
\right)
\label{eq:estimated_normalized_witness}
\end{equation}
is generally biased because the minimum eigenvalue and the minimum over orientations are nonlinear.  The bounds below therefore use a deterministic Lipschitz inequality, not unbiasedness of the witness.

For Hermitian $A$ and $B$, Weyl's inequality gives
\begin{equation*}
\left|\lambda_{\min}(A)-\lambda_{\min}(B)\right|
\leq\|A-B\|_{\rm op}.
\end{equation*}
Together with $|\min_\alpha a_\alpha-\min_\alpha b_\alpha|\leq\max_\alpha|a_\alpha-b_\alpha|$, $\|X\|_{\rm op}\leq\|X\|_{\rm HS}$, the isometry of $\operatorname{svec}$, and Eq.~\eqref{eq:B_normalization_contraction}, this yields
\begin{align}
\left|\widehat{\widetilde{\mathcal E}}-\widetilde{\mathcal E}\right|
&\leq
\max_{\alpha\in\{A,B\}}
\left\|
\widehat{\widetilde{\bar{\mathfrak M}}}_\alpha
-\widetilde{\bar{\mathfrak M}}_\alpha
\right\|_{\rm op}\nonumber\\
&\leq
\max_{\alpha\in\{A,B\}}
\frac{\left\|B_{\alpha,d}
(\widehat{\mathbf Y}_{N_U,N_S}-\mathbf Y)\right\|_2}
{C_\alpha (d_A,d_B)}\nonumber\\
&\leq\left\|\widehat{\mathbf Y}_{N_U,N_S}-\mathbf Y\right\|_2.
\label{eq:HS_norm_bound}
\end{align}
Because $\widehat{\mathbf Y}_{N_U,N_S}$ is unbiased,
\begin{equation}
\mathbb E\!\left[
\left\|\widehat{\mathbf Y}_{N_U,N_S}-\mathbf Y\right\|_2^2
\right]
=\operatorname{Tr}\operatorname{Cov}(\widehat{\mathbf Y}_{N_U,N_S}).
\label{eq:vector_mse_covariance}
\end{equation}
Markov's inequality applied to this squared Euclidean norm therefore gives the vector Chebyshev bound
\begin{align}
\mathbb P\!\left(
\left|\widehat{\widetilde{\mathcal E}}-\widetilde{\mathcal E}\right|>\epsilon
\right)
&\leq
\mathbb P\!\left(
\left\|\widehat{\mathbf Y}_{N_U,N_S}-\mathbf Y\right\|_2>\epsilon
\right)\nonumber\\
&\leq
\frac{\operatorname{Tr}\operatorname{Cov}(\widehat{\mathbf Y}_{N_U,N_S})}{\epsilon^2}.
\label{eq:app_chebyshev_bound}
\end{align}
Therefore, it suffices to control $\operatorname{Tr}\operatorname{Cov}(\widehat{\mathbf Y}_{N_U,N_S})$.

Equations~\eqref{eq:class_variables}--\eqref{eq:app_global_estimator_general} give the setting-wise and global estimators without treating the measurement outcomes as deterministic functions of the sampled unitaries.  Since the classes partition $\Omega^3$, each $\widehat{\mathbf Y}_{N_S}^{(s)}$ lies in the probability simplex.

Conditioned on all settings, setting-wise unbiasedness and conditional independence give
\begin{equation}
\mathbb E_\rho\!\left[
\widehat{\mathbf Y}_{N_U,N_S}\mid\mathbf U
\right]
=\frac1{N_U}\sum_{s=1}^{N_U}\mathbf Y(U_s),
\label{eq:app_cond_mean_global}
\end{equation}
and
\begin{equation}
\operatorname{Cov}\!\left(
\widehat{\mathbf Y}_{N_U,N_S}\mid\mathbf U
\right)
=\frac1{N_U^2}\sum_{s=1}^{N_U}
\operatorname{Cov}\!\left(
\widehat{\mathbf Y}_{N_S}^{(s)}\mid U_s
\right).
\label{eq:app_cond_cov_global}
\end{equation}
$
\operatorname{Cov}\!\left(
\widehat{\mathbf Y}_{N_U,N_S}\mid\mathbf U=\mathbf u
\right)
$
is the covariance of the global random vector \(\widehat{\mathbf Y}_{N_U,N_S}\), computed over all outcomes
$
\{I_{s,r}:s=1,\ldots,N_U,\ r=1,\ldots,N_S\}
$
under their conditional joint distribution given \(\mathbf U=\mathbf u\). 
\footnote{Explicitly,
$
\operatorname{Cov}\!\left(
\widehat{\mathbf Y}_{N_U,N_S}\mid\mathbf U=\mathbf u
\right)=
\mathbb E_\rho\!\left[
\mathbf J\mathbf J^{\mathsf T}
\,\middle|\,\mathbf U=\mathbf u
\right]
$  with $\mathbf J=\widehat{\mathbf Y}_{N_U,N_S}
-\frac1{N_U}\sum_{s=1}^{N_U}\mathbf Y(u_s)$. 
On the right,
$
\operatorname{Cov}\!\left(
\widehat{\mathbf Y}_{N_S}^{(s)}\mid U_s=u_s
\right)
$
is the covariance of the setting-specific random vector \(\widehat{\mathbf Y}_{N_S}^{(s)}\), computed only over the outcomes
$
I_{s,1},\ldots,I_{s,N_S}
$
drawn independently from \(p_\rho(\,\cdot\mid u_s)\).
Explicitly,
$
\operatorname{Cov}\!\left(
\widehat{\mathbf Y}_{N_S}^{(s)}\mid U_s=u_s
\right)=
\mathbb E_\rho\!\left[
\left(
\widehat{\mathbf Y}_{N_S}^{(s)}-\mathbf Y(u_s)
\right)
\left(
\widehat{\mathbf Y}_{N_S}^{(s)}-\mathbf Y(u_s)
\right)^{\mathsf T}
\,\middle|\,U_s=u_s
\right]$.
}
The equality follows because, conditioned on \(\mathbf U=\mathbf u\), the setting-specific estimators are independent, so their conditional cross-covariances vanish.
The law of total covariance is therefore
\begin{align}
\operatorname{Cov}\!\left(
\widehat{\mathbf Y}_{N_U,N_S}
\right)
&=
\mathbb E_{\mathbf U}\!\left[
\operatorname{Cov}_\rho \!\left(
\widehat{\mathbf Y}_{N_U,N_S}\mid\mathbf U
\right)
\right]\nonumber\\
&\quad+
\operatorname{Cov}_{\mathbf U}\!\left(
\mathbb E_\rho \!\left[
\widehat{\mathbf Y}_{N_U,N_S}\mid\mathbf U
\right]
\right),
\label{eq:app_total_cov_global}
\end{align}
where \(\mathbb E_{\mathbf U}\) and
\(\operatorname{Cov}_{\mathbf U}\) refer to the product Haar law of
\(\mathbf U=(U_1,\ldots,U_{N_U})\).  Substitution of
Eqs.~\eqref{eq:app_cond_mean_global} and
\eqref{eq:app_cond_cov_global} gives
\begin{align}
\operatorname{Cov}\!\left(
\widehat{\mathbf Y}_{N_U,N_S}
\right)
&=
\mathbb E_{\mathbf U}\!\left[
\frac1{N_U^2}\sum_{s=1}^{N_U}
\operatorname{Cov}\!\left(
\widehat{\mathbf Y}_{N_S}^{(s)}\mid U_s
\right)
\right]\nonumber\\
&\quad+
\operatorname{Cov}_{\mathbf U}\!\left(
\frac1{N_U}\sum_{s=1}^{N_U}\mathbf Y(U_s)
\right).
\label{eq:app_total_cov_global_explicit}
\end{align}
Because the settings \(U_1,\ldots,U_{N_U}\) are independent and
identically distributed, this simplifies to
\begin{align}
\operatorname{Cov}\!\left(
\widehat{\mathbf Y}_{N_U,N_S}
\right)
=
\frac1{N_U}
\mathbb E_{U_1}\!&\left[
\operatorname{Cov}_\rho \!\left(
\widehat{\mathbf Y}_{N_S}^{(1)}\mid U_1
\right)
\right]\nonumber\\
&+
\frac1{N_U}
\operatorname{Cov}_{U_1}\!\left(
\mathbf Y(U_1)
\right).
\label{eq:app_total_cov_global_iid}
\end{align}
For a fixed realization \(U_1=u\), the conditional covariance in the
first term of Eq.~\eqref{eq:app_total_cov_global_iid} is
\begin{align}
\operatorname{Cov}_\rho&\!\left(
\widehat{\mathbf Y}_{N_S}^{(1)}
\mid U_1=u
\right)
=\nonumber\\&
\mathbb E_\rho\!\left[
\left(
\widehat{\mathbf Y}_{N_S}^{(1)}-\mathbf Y(u)
\right)
\left(
\widehat{\mathbf Y}_{N_S}^{(1)}-\mathbf Y(u)
\right)^{\mathsf T}
\,\middle|\,U_1=u
\right].
\label{eq:app_single_setting_covariance_explicit}
\end{align}

By contrast, the covariance in the second term is taken only over the
random unitary setting:
\begin{align}
\operatorname{Cov}_{U_1}\!\left(
\mathbf Y(U_1)
\right)
&=
\mathbb E_{U_1}\!\left[
\left(
\mathbf Y(U_1)-\mathbf Y
\right)
\left(
\mathbf Y(U_1)-\mathbf Y
\right)^{\mathsf T}
\right],
\label{eq:app_unitary_covariance_explicit}
\end{align}
where
\(\mathbf Y=\mathbb E_{U_1}[\mathbf Y(U_1)]\).
No measurement-outcome randomness remains in this covariance because
\(\mathbf Y(u)\) is deterministic for every fixed \(u\).

Thus, the first term in
Eq.~\eqref{eq:app_total_cov_global_iid} is the Haar average of the
measurement-outcome covariance at a fixed unitary and represents shot
noise.  The second term is the covariance, over Haar-random settings,
of the conditional class-probability vector and represents the
fluctuation caused by sampling only finitely many random unitaries.

Every component of the simplex-valued vector $\widehat{\mathbf Y}_{N_S}^{(1)}$ lies in $[0,1]$, so Appendix~\ref{App:max_variance} gives
\begin{equation}
\operatorname{Tr}\operatorname{Cov}(\widehat{\mathbf Y}_{N_S}^{(1)}\mid U_1)
=\sum_{\mu=0}^9\operatorname{Var}([\widehat{\mathbf Y}_{N_S}^{(1)}]_\mu\mid U_1)
\leq\frac{10}{4}.
\end{equation}
Likewise, $\mathbf Y(U)$ is simplex-valued and $\operatorname{Tr}\operatorname{Cov}_U(\mathbf Y(U))\leq10/4$.  Hence
\begin{equation}
\operatorname{Tr}\operatorname{Cov}(\widehat{\mathbf Y}_{N_U,N_S})
\leq\frac5{N_U}.
\label{eq:coarse_trace_covariance}
\end{equation}
For a fixed total number $N_{\rm tot}=N_UN_S$ of measurement shots, this coarse bound is smallest at $N_S=3$.  In that case each setting contributes the single one-hot vector $\mathbf Z_{s,123}$, and Eq.~\eqref{eq:app_chebyshev_bound} give

\begin{equation}
\mathbb P_\rho\!\left(
\left|\widehat{\widetilde{\mathcal E}}-\widetilde{\mathcal E}\right|>\epsilon
\right)
\leq\frac{15}{N_{\rm tot}\epsilon^2}.
\label{eq:coarse_chebyshev_witness}
\end{equation}
Thus an error guarantee with tolerance $\epsilon$ and failure probability at most $\delta$ follows from
\begin{equation}
N_{\rm tot}\geq\frac{15}{\epsilon^2\delta}=\frac{3|\{\mathcal C _\mu\}_{\mu}|}{2\epsilon^2\delta}.
\label{eq:app_B.H.C._Ntot}
\end{equation}
If the true normalized witness is negative, 
its sign margin is $\gamma:=|\widetilde{\mathcal E}(\rho)|$.  The separate sufficient sign guarantee is $N_{\rm tot}\geq15/(\gamma^2\delta)$.  A directly checkable observed-data certificate is
\begin{equation}
N_{\rm tot}\geq\frac{15}{2\delta} \widehat{\widetilde{\mathcal E}}^{-2}.
\label{eq:chebyshev_observed_certificate}
\end{equation}
For comparison, full state tomography of $\rho$ requires $\sim D^3\delta^{-1}\epsilon^{-2}$ measurements to achieve entrywise tolerance $\epsilon$ with failure probability at most $\delta$ where $D=d_A d_B$.  Weyl's inequality applied after partial transposition gives
\begin{equation*}
\left|\lambda_{\min}(\widehat\rho^\Gamma)-\lambda_{\min}(\rho^\Gamma)\right|
\leq\|\widehat\rho^\Gamma-\rho^\Gamma\|_{\rm HS}
=\|\widehat\rho-\rho\|_{\rm HS}.
\end{equation*}
Thus it is enough to control the Hilbert--Schmidt error, whose scaling is $\sim D^2\delta^{-1}\epsilon^{-2}$, leading to \cite{Vermersch2024}
%
\begin{equation}
    N_{\rm PPT} \sim \frac{D^2}{\delta}[\lambda_{\rm min}(\rho^\Gamma)]^{-2}
\end{equation}
At fixed normalized tolerance and confidence, our bound for $N_{\rm tot}$ has no explicit dimension factor, whereas $N_{\rm PPT}$ has a quadratic factor; sign recovery in either method still depends on its state-dependent true margin respectively $\widehat{\widetilde{\mathcal{E}}}$ and $\lambda_{\rm min}$ and cannot be upper-bounded; indeed, $\widetilde{\mathcal E}$ (or $\lambda_{\rm min}(\rho^\Gamma)$) can be arbitrarily close to zero, and then assuring its sign can require arbitrarily big number of measurements.

The true normalized margin can itself depend on dimension.  For $d_A < d_B$ and a maximally entangled state,
\begin{align}
\lambda_{\min}\!\bigl((\proj{\Phi_{d_A,d_B}})^{T_A}\bigr)
&=-\frac1{d_A},\label{eq:comparison_minimal_eigenvalues}\\
\widetilde{\mathcal E}(\proj{\Phi_{d_A,d_B}})
&\sim
-\frac1{2d_A}
\nonumber
\end{align}
(see Appendix \ref{sm:S3} for derivation). Hence
\begin{align}
    N_{\rm tot} \sim 4 d_A^2 \delta^{-1}, \quad
    N_{\rm PPT} \sim D^2 d_A^2 \delta^{-1}. \nonumber
\end{align}
Our certification gains an additional dimension-dependent factor, but the same factor (up to a scalar) is also gained in the $PPT$ certification.
This is the main advantage that we found in this work: a finite-size certification theorem governed only by low-order invariant statistics that avoids the parameter-count overhead of full state tomography.

Furthermore, a separate exponentially concentrated estimate follows by applying McDiarmid's inequality to $\|\widehat{\mathbf Y}_{N_U,N_S}-\mathbf Y\|_2$.  As shown in Appendix~\ref{app:concentration_ustat}, for the minimal-triple choice $N_S=3$, an error tolerance $\epsilon$ with failure probability at most $\delta$ is guaranteed by
\begin{equation}
N_{\rm tot}\geq
\frac3{\epsilon^2}
\left(1+\sqrt{\ln(1/\delta)}\right)^2.
\label{eq:main_McDiarmid_tolerance}
\end{equation}
The requirement of $\epsilon<\widehat{\widetilde{\mathcal E}}$ yields the observed-data certificate
\begin{equation}
N_{\rm tot}\geq
3
\left(1+\sqrt{\ln(1/\delta)}\right)^2
\widehat{\widetilde{\mathcal E}}^{-2}.
\end{equation}
Crucially, this bound has no explicit dependence on the number of equivalence classes; at fixed total shots, its minimal-triple specialization has the factor $N_S=3$, linear in the order of the correlator.

\section{Performance}
The matrix \eqref{eq:sm_MbarExplicit_correct} is driven by differences between global and marginal moments (e.g.\ $x_3-x_5=\Tr\rho_A^2-\Tr\rho^2$ and $x_8-x_S$), so it is strongest when the state is globally pure but locally mixed. Accordingly, the separability criterion is most sensitive near maximally entangled states, where the global state is nearly pure but the marginals are nearly maximally mixed.
For $d_A=d_B$ and $\rho=\proj{\Phi_{d_A}}$ with $\ket{\Phi_{d_A}}=\frac1{\sqrt d_A}\sum_{j=0}^{d_A-1}\ket{jj}$, we obtain
\begin{align}
\lambda_{\min}\!\big(\bar{\mathfrak{M}}(\proj{\Phi_{d_A}})\big)
=&(d_A-1)\,
\frac{2d_A^2-d_A+5-\sqrt{s}}{4d_A^2},\label{eq:MESformula}\\
s=&\ 4d_A^4+4d_A^3+29d_A^2+6d_A+41.\nonumber
\end{align}
As $d_A\to\infty$, the maximal violation approaches $-1/2$, setting the natural high-dimensional scale of the separability criterion.
To quantify robustness against white noise, we next consider the isotropic family
\begin{equation}
\rho_{\rm iso}(p)=p\,\proj{\Phi_{d_A}}+(1-p)\frac{\id_{d_A^2}}{d_A^2},\qquad p\in[0,1].
\label{eq:isodef}
\end{equation}
This family is a sharp benchmark because, for isotropic states, PPT, separability, and positivity under the reduction map coincide at
\begin{equation}
p_{\rm PPT}(d)=\frac{1}{d_A+1}\sim d_A^{-1}.
\label{eq:pPPT}
\end{equation}
The corresponding second-order benchmark is the purity inequality $\Tr(\rho^2)\le \Tr(\rho_A^2)$ inherited directly from the reduction criterion. It detects entanglement only for
\begin{equation}
p>p_{\rm purity}(d_A):=\frac{1}{\sqrt{d_A+1}}\sim d_A^{-1/2}.
\label{eq:ppurity}
\end{equation}
Our third-order witness instead detects entanglement already at (see App. \ref{sm:S3})
\begin{equation}
p>p_{\rm 3rd}(d_A):=\frac{4-d_A^2+d_A\sqrt{d_A^2+8}}{4(d_A+1)}
= \frac{2}{d_A}+O(d_A^{-2}).
\label{eq:p3rd}
\end{equation}

\subsection{Why the $V_0=\mathbf I$ direction matters}
Isotropic states expose the asymptotic scaling but hide the role of the identity direction, since there $\rho_A\otimes\id$ and $\id\otimes\rho_B$ are proportional to $\id$.
To isolate its effect, consider the biased two-qubit family instead
\begin{equation}
\rho_{x,p}:=(1-p)\proj{00}+p\,\proj{\psi_x},
\label{eq:Xfamily}
\end{equation}
where $\ket{\psi_x}:=\sqrt{x}\ket{01}+\sqrt{1-x}\ket{10}$ for $x\in[0,1]$. 
Here the local marginals are biased, so the marginal sectors span directions independent of the identity.
For this family, the full  $\bar{\mathfrak M}(\rho_{x,p})$ and the homogeneous $3\times3$ block obtained by dropping the identity sector can be evaluated exactly.
A direct calculation shows that $\det\bar{\mathfrak M}(\rho_{x,p})\propto(p-1)(2p-1)$ up to non-negative factors, so the four-direction criterion detects entanglement for all $p>p_{\rm aff}=1/2$,
independently of $x$, whereas the homogeneous block turns non-PSD only at larger thresholds; for example, $p_{\rm hom}(x=0.5)\approx0.608$, with similarly higher values across representative $x$.
Thus the affine extension strictly enlarges the detectable region on a nonisotropic family, and the gain originates from the additional operator directions supplied by the marginals.
The family also exposes the intrinsic asymmetry of the reduction map: because $\mathfrak{R}=\rho_A\otimes\id-\rho$, the thresholds are not symmetric under $x\to1-x$, and exchanging the reduced subsystem interchanges them.
This asymmetry confirms that the witness probes the marginal sectors, not merely the global purity.

\section{Comparison with the related works}
%
%

During the work on this article, another interesting result creating entanglement criteria using third-order local invariants appeared in Ref.~\cite{Tang2026WitnessExpansion}. 

For an entanglement witness $W$ the authors define
\begin{equation*}
    \widetilde{W} = I - W / \lVert W \rVert_{\rm sup},\quad \Tr(\widetilde W \rho) > 1 \Longrightarrow \rho \text{ is entangled}.
\end{equation*}
Positivity of the modified witness lets to define a norm:
\begin{equation*}
    \|\rho \|_{\widetilde{W},\mathcal{U}(d_A,d_B),3}^3:= \Tr \left[ \mathbb E_{U}  \left(U^{\dagger \otimes 3} \widetilde{W}^{\otimes 3} U^{\otimes 3}  \right) \rho^{\otimes 3} \right],
\end{equation*}
being expected value of $\widetilde W$ averaged over local unitaries. 
If the averaged expected value exceeds $1$, it exceeds it for some $U$, proving the state $\rho$ is entangled. 

Observe that the expression can be obtained applying a local Haar-twirling superoperator:
\begin{equation*}
    \mathcal{T}_{3} (X) = \mathbb{E}_U U^{\dagger \otimes 3} X U^{\otimes 3}, \quad U \in U(d_A) \otimes U(d_B)
\end{equation*}
to the left factor of the HS inner product:
\begin{equation*}
    \Tr (\widetilde W \rho) ^ 3 = \Tr (\widetilde W^{\otimes 3} \rho^{\otimes 3}) = \langle \widetilde W^{\otimes 3} | \rho^{\otimes 3} \rangle_{HS} 
\end{equation*}

We can understand the above expression as:
\begin{equation*}
    \Tr \left[ \mathbb E_{U}  \left(U^{\dagger \otimes 3} \widetilde{W}^{\otimes 3} U^{\otimes 3}  \right) \rho^{\otimes 3} \right] = \langle \mathcal{T}_3 \widetilde W^{\otimes 3}| \rho^{\otimes n}\rangle_{HS}
\end{equation*}
As the Haar-twirling superoperator $\mathcal{T}_3$ is a self-adjoint projector in the operator space, it can be rewritten as:
\begin{equation*}
    \langle \mathcal{T}_3 \widetilde W^{\otimes 3}| \mathcal{T}_3\rho^{\otimes 3}\rangle_{HS} = \sum_{i} \alpha_i y_i x_i   
\end{equation*}
where $\vec x$ is the third-order local invariant vector of $\rho$ and $\vec y$ is the third-order local invariant vector of $\widetilde W$. A given $W$ produces an affine condition on the vector of local invariants related to $\rho$.

In contrast, our criteria maps linearly this vector to a $4\times 4$ matrix, which is then examined to be PSD, leading to a criterion in  Eq. \eqref{eq:Esym_def} that is non-linear in the entries of the vector of local invariants.

\section{Conclusions}

We converted the reduction criterion into a measurable third-order separability test.
Testing $\mathfrak{R}_{A/B}(\rho)$ exactly requires access to the full operator space and is therefore tomographically hard.
Our construction instead restricts the quadratic form to the four-dimensional subspace spanned by $\{\id,\rho_A\otimes\id_B,\id_A\otimes\rho_B,\rho\}$, yielding for each orientation a measurable 4 × 4 matrix; combined via direct sum as in Eq. \eqref{eq:Esym_def}, these give an 8 × 8 matrix built from a small set of local invariants, whose minimum eigenvalue defines the separability criterion 
$\mathcal{E}(\rho)$.
Its minimum eigenvalue furnishes an entanglement certificate whose error bounds have no explicit local-dimension dependence at fixed normalized tolerance and confidence; sign certification additionally depends on the true normalized margin. 
Operationally, the method interpolates between simple purity witnesses and fully tomographic separability tests.

For isotropic states, third-order information improves the detection threshold from $d^{-1/2}$ to $O(d^{-1})$, matching the scaling of the PPT/separability boundary.
Because the construction is based on the reduction criterion, it inherits the same distillation-oriented physical motivation \cite{Yi2026}.
For isotropic states, where PPT, reduction, and separability coincide, this separability criterion approaches the distillability boundary.
On the non-isotropic benchmark, third-order information strictly enlarges the certification region.

\section*{Acknowledgments}
We thank C.~Lupo, D.~Chru\'{s}ci\'{n}ski and M. Ziman for several insights.
GSc is supported by grant NCN SONATA 21 UMO2025/59/D/ST2/02708.
GSc thanks the Institute of Physics at the Slovak Academy of Sciences for its hospitality through the SAIA Support Fund.

\bibliographystyle{apsrev4-2}

\bibliography{rand_v11bib}

\appendix
\section{Explicit linear reconstruction of the moment matrices}
\label{app:explicit_linear_maps}

In this appendix, we give the explicit linear maps appearing in
Eq.~\eqref{eq:mvec}. 

The two vectorizations of $\bar{\mathfrak{M}}_{A/B}$ can be written directly as
\begin{equation}
\mathbf m_A=
\begin{pmatrix}
d_B-1\\
\sqrt2(d_B-1)x_3\\
\sqrt2(1-x_1)\\
\sqrt2(x_3-x_5)\\
(d_B-1)x_7\\
\sqrt2(x_3-x_4)\\
\sqrt2(x_7-x_8)\\
x_1-x_2\\
\sqrt2(x_4-x_6)\\
x_8-\frac12 x_S
\end{pmatrix},
\qquad
\mathbf m_B=
\begin{pmatrix}
d_A-1\\
\sqrt2(d_A-1)x_1\\
\sqrt2(1-x_3)\\
\sqrt2(x_1-x_5)\\
(d_A-1)x_2\\
\sqrt2(x_1-x_4)\\
\sqrt2(x_2-x_6)\\
x_3-x_7\\
\sqrt2(x_4-x_8)\\
x_6-\frac12 x_S
\end{pmatrix}.
\label{eq:app_m_vectors}
\end{equation}

From explicit moment matrices in the main text, one obtains by linearity,
\begin{widetext}
\begin{equation}
A_{\bar{\mathfrak M}_A}
=
\begin{pmatrix}
0&0&0&0&0&0&0&0&0\\
0&0&\sqrt2(d_B-1)&0&0&0&0&0&0\\
-\sqrt2&0&0&0&0&0&0&0&0\\
0&0&\sqrt2&0&-\sqrt2&0&0&0&0\\
0&0&0&0&0&0&d_B-1&0&0\\
0&0&\sqrt2&-\sqrt2&0&0&0&0&0\\
0&0&0&0&0&0&\sqrt2&-\sqrt2&0\\
1&-1&0&0&0&0&0&0&0\\
0&0&0&\sqrt2&0&-\sqrt2&0&0&0\\
0&0&0&0&0&0&0&1&-\frac12
\end{pmatrix},
\qquad
\mathbf b_A=
\begin{pmatrix}
d_B-1\\0\\\sqrt2\\0\\0\\0\\0\\0\\0\\0
\end{pmatrix},
\label{eq:app_A_A}
\end{equation}
\end{widetext}
and analogously for $A_{\bar{\mathfrak M}_B}$. Thus,
\begin{equation}
A_{\bar{\mathfrak M}_{A/B}}\in\mathbb R^{10\times9},
\qquad
\mathbf b_{A/B}\in\mathbb R^{10}.
\end{equation}
To connect Eq.~\eqref{eq:mvec} to the randomized-measurement
data, introduce the full invariant vector
\begin{equation}
\mathbf x_{\mathrm{full}}
=(x_0,x_1,\ldots,x_8,x_S)^{\mathsf T}\in\mathbb R^{10},
\label{eq:app_x_full}
\end{equation}
with $x_0=(\Tr\rho)^3=1$.
The full vector $\mathbf x_{\mathrm{full}}$ is used only in the forward Weingarten relation.  After applying the projection $P$, the fixed component $x_0=1$ is not treated as an independent invariant; its constant contribution to the affine moment-matrix maps is collected in $\mathbf b_{A/B}$.

Let $\mathbf y=(y_0,\ldots,y_9)^{\mathsf T}\in\mathbb R^{10}$
be the Haar-averaged correlator vector introduced in Sec. \ref{sec:Randomized_measurements}. The third-order
Weingarten calculation gives~\cite{QST2025}
\begin{equation}
\mathbf y=M(d_A,d_B)\mathbf x_{\mathrm{full}}.
\label{eq:app_forward_reconstruction}
\end{equation}
where
\begin{equation}
    M(d_A,d_B)=\omega_A\omega_B\, D(d_A,d_B)\, R(d_A,d_B),
\label{eq:app_M_factorization}
\end{equation}
and 
\begin{equation}
\omega_\alpha=
\frac{1}{d_\alpha(d_\alpha^2-1)(d_\alpha^2-4)},
\qquad \alpha\in\{A,B\}.
\end{equation}
The diagonal factor is
\begin{equation}
 D(d_A,d_B)
=
\operatorname{diag}
\begin{pmatrix}
1,\\
d_B-2,\\
(d_B-1)(d_B-2),\\
d_A-2,\\
(d_A-2)(d_B-2),\\
(d_A-2)(d_B-2),\\
(d_A-2)(d_B-1)(d_B-2),\\
(d_A-1)(d_A-2),\\
(d_A-1)(d_A-2)(d_B-2),\\
(d_A-1)(d_A-2)(d_B-1)(d_B-2)
\end{pmatrix}.
\label{eq:app_D_matrix}
\end{equation}
In the column ordering of Eq.~\eqref{eq:app_x_full}, the remaining
factor is
\begin{widetext}
\begin{equation}
\resizebox{\textwidth}{!}{$
 R(d_A,d_B)=
\begin{pmatrix}
(d_A^2-2)(d_B^2-2)
&3d_B(2-d_A^2)&4d_A^2-8&3d_A(2-d_B^2)
&6d_Ad_B&3d_Ad_B&-12d_A&4d_B^2-8&-12d_B&16\\
(d_A^2-2)(d_B+1)
&(d_A^2-2)(d_B-1)&4-2d_A^2&-3d_A(d_B+1)
&2d_A(1-d_B)&d_A(1-d_B)&6d_A&4d_B+4&4d_B-4&-8\\
d_A^2-2
&3d_A^2-6&2d_A^2-4&-3d_A
&-6d_A&-3d_A&-6d_A&4&12&8\\
(d_A+1)(d_B^2-2)
&-3d_B(d_A+1)&4d_A+4&(d_A-1)(d_B^2-2)
&2d_B(1-d_A)&d_B(1-d_A)&4d_A-4&4-2d_B^2&6d_B&-8\\
d_Ad_B+d_A+d_B+1
&d_Ad_B-d_A+d_B-1&-2d_A-2&d_Ad_B+d_A-d_B-1
&d_Ad_B+2&-d_A-d_B-1&2-2d_A&-2d_B-2&2-2d_B&4\\
d_Ad_B+d_A+d_B+1
&d_Ad_B-d_A+d_B-1&-2d_A-2&d_Ad_B+d_A-d_B-1
&-2d_A-2d_B-2&d_Ad_B+d_A+d_B+3&2-2d_A&-2d_B-2&2-2d_B&4\\
d_A+1
&3d_A+3&2d_A+2&d_A-1
&2d_A-2&d_A-1&2d_A-2&-2&-6&-4\\
d_B^2-2
&-3d_B&4&3d_B^2-6
&-6d_B&-3d_B&12&2d_B^2-4&-6d_B&8\\
d_B+1
&d_B-1&-2&3d_B+3
&2d_B-2&d_B-1&-6&2d_B+2&2d_B-2&-4\\
1&3&2&3&6&3&6&2&6&4
\end{pmatrix}.
$}
\label{eq:app_R_matrix}
\end{equation}
\end{widetext}
The last column in Eq.~\eqref{eq:app_R_matrix} is obtained by removing one of the two identical columns multiplying $x_9$ and $x_{10}$ in the
original Weingarten system and introducing
$x_S=x_9+x_{10}$.
Let
\begin{equation}
P=
\begin{pmatrix}
0_{9\times1}&\mathbb I_9
\end{pmatrix}
\in\mathbb R^{9\times10},
\label{eq:app_projection_P}
\end{equation}
the map that removes the normalization component $x_0$. 

Therefore, since the matrix $M(d_A,d_B)$ is invertible,
\begin{equation*}
\mathbf x
=P\,M(d_A,d_B)^{-1}\mathbf y
=:L(d_A,d_B)\mathbf y,
\;
L(d_A,d_B)\in\mathbb R^{9\times10}.
\end{equation*}
If instead one works with the probability vector
$\mathbf Y=K\mathbf y$, then
\begin{equation}
\mathbf x=L(d_A,d_B)K^{-1}\mathbf Y.
\label{eq:app_L_from_Y}
\end{equation}
Combining Eqs.~\eqref{eq:mvec} and
\eqref{eq:app_L_from_Y} gives
\begin{equation}
\mathbf m_\alpha
=A_{\bar{\mathfrak M}_\alpha}
L(d_A,d_B)K^{-1}\mathbf Y+\mathbf b_\alpha,
\qquad \alpha\in\{A,B\}.
\label{eq:app_complete_map}
\end{equation}
Hence the complete reconstruction map is
\begin{equation}
B_{\alpha,d}
=A_{\bar{\mathfrak M}_\alpha}
L(d_A,d_B)K^{-1}
\in\mathbb R^{10\times10},
\qquad \alpha\in\{A,B\}.
\label{eq:app_B_map}
\end{equation}
\section{Maximally entangled benchmark: derivation of Eq.~\eqref{eq:MESformula} in the main text}
\label{sm:S3}

This section derives the closed-form minimum eigenvalue $\lambda_{\min}(\bar{\mathfrak{M}_{A/B}}(\proj{\Phi_{d_A,d_B}}))$.
Let
$d_A \le d_B$
and
$\rho=\proj{\Phi_{d_A,d_B}}$
with
$\ket{\Phi_{d_A,d_B}}=\frac1{\sqrt d_A}\sum_{j=0}^{d_A-1}\ket{jj}$.
Then $\rho$ is pure and maximally entangled, so
$\rho_A = \id/d_A$, $\rho_B = P_{\{e_1, \dots, e_{d_A}\}}/d_A$.
A direct calculation gives
\begin{align}
x_1&=x_3=\Tr\Big(\Big(\frac{\id}{d_A}\Big)^2\Big)=\frac{1}{d_A}, \nonumber\\
x_2&=x_7=\Tr\Big(\Big(\frac{\id}{d_A}\Big)^3\Big)=\frac{1}{d_A^2},\nonumber\\
x_5&=\Tr(\rho^2)=1,\\
x_4&=\Tr\big[(\rho_A\otimes\rho_B)\rho\big]=\Tr\Big(\frac{\id}{d_A^2}\rho\Big)=\frac{1}{d_A^2},\nonumber\\
x_6&=x_8=\Tr\big[\rho_B\,\Tr_A(\rho^2)\big]=\Tr\big[\rho_B^2\big]=\frac1{d_A} \nonumber.
\end{align}
Moreover, $(\proj{\Phi_{d_A,d_B}})^{T_A}=\frac1{d_A}\,F$, where $F$ is the swap operator,
so $F^3=F$ and $\Tr F=d$, implying $\Tr[(\rho^{T_A})^3]=1/{d_A}^2$.
Hence
\begin{equation}
x_S=\frac12\Big(1+\frac1{d_A^2}\Big).
\end{equation}
Substituting these invariants into \eqref{eq:sm_MbarExplicit_correct} yields:
\begin{align}
    \bar{\mathfrak{M}}_A 
    &= 
    \begin{pmatrix}
        d_A-1 & \frac{d_A-1}{d_A}&\frac{d_A-1}{d_A}&-\frac{d_A-1}{d_A}\\
        \cdot & \frac{d_A-1}{d_A^2}&\frac{d_A-1}{d_A^2}&-\frac{d_A-1}{d_A^2}\\
        \cdot & \cdot & \frac{d_A-1}{d_A^2}&-\frac{d_A-1}{d_A^2}\\
        \cdot &\cdot &\cdot &-\frac{(d_A-1)^2}{2d_A^2}
    \end{pmatrix},
    \\
    \bar{\mathfrak{M}}_B 
    &= 
    \begin{pmatrix}
        d_B-1 & \frac{d_B-1}{d_A}&\frac{d_B-1}{d_A}&-\frac{d_A-1}{d_A}\\
        \cdot & \frac{d_B-1}{d_A^2}&\frac{d_A-1}{d_A^2}&-\frac{d_A-1}{d_A^2}\\
        \cdot & \cdot & \frac{d_A-1}{d_A^2}&-\frac{d_A-1}{d_A^2}\\
        \cdot &\cdot &\cdot &-\frac{(d_A-1)^2}{2d_A^2}
    \end{pmatrix}
    \\
    &= 
    (d_B-d_A)
    \begin{pmatrix}
        1 & \frac1{d_A}& 0 & 0\\
        \cdot & \frac1{d_A^2} & 0 & 0 \\
        \cdot & \cdot &  0 & 0 \\
        \cdot &\cdot &\cdot & 0
    \end{pmatrix} + \bar{\mathfrak{M}}_A
\end{align}
Hence $\bar{\mathfrak{M}}_B \succeq \bar{\mathfrak{M}}_A$ and it is enough to consider the sign of minimal eigenvalue of 
\begin{align}
    \bar{\mathfrak{M}}_A(\proj{\Phi_{d_A,d_B}})
    = (d_A - 1)
    \left[ \begin{array}{l} 
    1 \\ \frac 1{d_A} \\ \frac 1{d_A} \\ -\frac1{d_A} 
    \end{array} \right] \times \mathrm{h.c.}
    \nonumber \\
    - \frac{d_A^2-1}{2d_A^2} 
    \left[ \begin{array}{l} 
    0 \\ 0 \\ 0 \\ 1 
    \end{array} \right] \times \mathrm{h.c.}
    =: \ket{\alpha}\bra{\alpha} - \ket{\beta}\bra{\beta}.
\end{align}
Its eigenvalues are:
\begin{align}
\lambda_\pm
=&
\frac{
\langle\alpha|\alpha\rangle
-
\langle\beta|\beta\rangle
\pm
\sqrt{
\left(
\langle\alpha|\alpha\rangle
+
\langle\beta|\beta\rangle
\right)^2
-
4|\langle\alpha|\beta\rangle|^2
}
}{2} 
\nonumber \\
=&(d_A-1)\,
\frac{2d_A^2-d_A+5\pm\sqrt{s}}{4d_A^2},
\end{align}
where $s = 4d_A^4+4d_A^3+29d_A^2+6d_A+41$.
One has:
\begin{align}
    \widetilde{\mathcal E}(\proj{\Phi_d}) = \frac{\lambda_-}{\lambda_+}
    \nonumber \\
    \frac{2d_A^2-d_A+5-\sqrt{s}}{2d_A^2-d_A+5+\sqrt{s}}
\end{align}
which gives Eq. \eqref{eq:comparison_minimal_eigenvalues} in the main text. When $d_A = d_B = d$, for unnormalised matrix  $\bar{\mathfrak{M}}(\proj{\Phi_d}) = \lambda_-$ and we recover the formula~\eqref{eq:MESformula} in the main text.

For $d=2$ this gives
\begin{equation}
    \frac{11- \sqrt{265}}{16} \approx -0.329926
\end{equation}
In the case of equal local dimensions, define the isotropic family ($\Phi_{d_A}=\proj{\Phi_{d_A}}$)
\begin{equation}
\rho_{\rm iso}(p)=p\,\proj{\Phi_{d_A}}+(1-p)\frac{\id_{d_A^2}}{d_A^2},\qquad p\in[0,1].
\end{equation}
This family satisfies $\rho_A=\rho_B=\id/d$ for all $p$.
Consequently the operators $V_1=\rho_A\otimes\id$ and $V_2=\id\otimes\rho_B$ are both proportional to $V_0=\id$,
so the added identity direction does not generate new independent constraints; this is the sense in which the affine construction is redundant here.
\begin{align*}
    &x_1 = x_3  = \frac 1d_A, 
    \qquad x_2 = x_7 = x_4  = \frac 1{d_A^2} \\
    &x_5  = p^2 + \frac {1-p^2}{d_A^2},\qquad 
    x_8 = x_6  = \frac{x_5}{d_A}
\end{align*}
Finally:
\begin{align*}
    \mathrm{Tr} \rho^3 & = p^3 + 3 \frac{p^2(1-p)}{d_A^2} + 3 \frac{p(1-p)^2}{d_A^4} +\frac{(1-p)^3}{d_A^4} \\
    \mathrm{Tr} (\rho^{T_A})^3 & = \frac{p^3}{d_A^2} + 3 \frac{p^2(1-p)}{d_A^2} + 3 \frac{p(1-p)^2}{d_A^4} + \frac{(1-p)^3}{d_A^4}
\end{align*}
and hence:
\begin{align*}
    x_S = p^3 \frac{d_A^4 - 5 d_A^2 + 4}{2d_A^4} + 3 p^2 \frac{d_A^2 - 1}{d_A^4} + \frac 1{d_A^4}
\end{align*}
One has:
\begin{align*}
    \bar{\mathfrak{M}}(\rho)
    & =
    \frac{d_A-1}{d_A^2}
    uu^T
    +
    \left(x_8 - x_S - \frac 1{d_A-1} (x_3-x_5)^2 \right)
    vv^T,
\end{align*}
where $u=[d_A, 1, 1, (1-p^2(d+1))/d_A]$ and $v=[0,0,0,1]$.
The above is PSD iff:
\begin{align}
    x_8 - x_S - \frac 1{d_A-1} (x_3-x_5)^2 \ge 0
\end{align}
hence if a state is separable, then:
\begin{equation}
-\frac{(d_A-1)(d_A+1)p^2}{2d_A^4}\,\Big(2(d_A+1)p^2+(d_A^2-4)p-2(d_A-1)\Big) \ge 0.
\label{eq:factorization}
\end{equation}
For completeness, one may compute
\begin{equation}
\Tr(\rho_{\rm iso}^2)=p^2+\frac{1-p^2}{d_A^2},
\end{equation}
and all other invariants needed for $\bar{\mathfrak{M}}$ can be expressed as rational functions of $p$ and $d_A$.

Because of the proportionality relations among $V_0,V_1,V_2$, the $4\times4$ matrix
$\bar{\mathfrak{M}}(\rho_{\rm iso}(p))$ can be brought by a congruence transformation to a matrix with a $2\times2$ nontrivial block
plus two zero directions.
Positivity of $\bar{\mathfrak{M}}$ is therefore equivalent to positivity of that $2\times2$ block.
Evaluating the resulting block condition yields the quadratic inequality
\begin{equation}\label{eq:piso}
2(d_A+1)p^2+(d_A^2-4)p-2(d_A-1)\le 0,
\end{equation}
hence the violation (entanglement detection) occurs for
\begin{equation}
p>\frac{4-d_A^2+d_A\sqrt{d_A^2+8}}{4(d_A+1)}=\frac{4(d_A-1)}{d_A\sqrt{d_A^2+8}+d_A^2-4}\sim\frac2{d_A}
\end{equation}
For $d_A=3$ this gives $p\approx0.4606$, which is better than the threshold $p=1/\sqrt[3]{10}\approx0.4641$ of the third-order criterion in \cite{QST2025}.

\section{Maximal variance}
\label{App:max_variance}

We recall the elementary variance bound used in the main text.

\begin{proposition}
Let \(X\) be a real random variable satisfying \(0\leq X\leq1\) almost surely. Then
$\operatorname{Var}(X)\leq\frac14.$
Equality holds if and only if \(X\) is Bernoulli distributed with
$\mathbb P(X=0)=\mathbb P(X=1)=\frac12$.
In particular, for every event \(A\),
\begin{equation}
\operatorname{Var}(\mathbbm 1_A)
=
\mathbb P(A)\bigl[1-\mathbb P(A)\bigr]
\leq\frac14.
\end{equation}
\end{proposition}

\begin{proof}
Since \(0\leq X\leq1\), one has
\begin{equation}
\left(X-\frac12\right)^2\leq\frac14
\end{equation}
almost surely. Therefore,
\begin{align}
\operatorname{Var}(X)
&=
\mathbb E\!\left[
\left(X-\frac12\right)^2
\right]
-
\left(\mathbb E[X]-\frac12\right)^2
\nonumber\\
&\leq
\mathbb E\!\left[
\left(X-\frac12\right)^2
\right]
\leq\frac14.
\end{align}
Equality requires both
\(\mathbb E[X]=\tfrac12\) and
\(\left(X-\tfrac12\right)^2=\tfrac14\) almost surely. The latter condition implies \(X\in\{0,1\}\) almost surely, while the former then implies
\(\mathbb P(X=0)=\mathbb P(X=1)=\tfrac12\).
\end{proof}

\section{Derivation of McDiarmid inequality - based bound}
\label{app:concentration_ustat}
For the setting-block McDiarmid argument, define the independent setting-outcomes random variables:
\begin{equation}
W_s:=(U_s,I_{s,1},\ldots,I_{s,N_S}),
\qquad s=1,\ldots,N_U,
\end{equation}
and let $F:=\|\widehat{\mathbf Y}_{N_U,N_S}-\mathbf Y\|_2$.  Since the setting-wise estimators are i.i.d., unbiased, and simplex-valued,
\begin{align}
\mathbb E_\rho[F^2]
&=\frac1{N_U}\,
\mathbb E_\rho\!\left[
\|\widehat{\mathbf Y}_{N_S}^{(1)}-\mathbf Y\|_2^2
\right]\nonumber\\
&=\frac1{N_U}\left(
\mathbb E_\rho\|\widehat{\mathbf Y}_{N_S}^{(1)}\|_2^2
-\|\mathbf Y\|_2^2
\right)
\leq\frac1{N_U}.
\label{eq:app_McDiarmid_mean_square}
\end{align}
Jensen's inequality therefore gives $\mathbb E_\rho[F]\leq N_U^{-1/2}$.  Replacing one random variable $W_s$ changes the associated simplex-valued setting estimator and hence changes $F$ by at most
\begin{equation}
|F-F'|\leq\frac1{N_U}
\|\widehat{\mathbf Y}_{N_S}^{(s)}
-\widehat{\mathbf Y}_{N_S}^{(s)\prime}\|_2
\leq\frac{\sqrt2}{N_U}.
\end{equation}
McDiarmid's inequality now yields
\begin{equation}
\mathbb P_\rho\!\left(F\geq\mathbb E_\rho[F]+t\right)
\leq\exp(-N_Ut^2).
\end{equation}
With $t=\sqrt{\ln(1/\delta)/N_U}$, we obtain
\begin{equation}
\mathbb P_\rho\!\left(
\|\widehat{\mathbf Y}_{N_U,N_S}-\mathbf Y\|_2
\geq\frac{1+\sqrt{\ln(1/\delta)}}{\sqrt{N_U}}
\right)\leq\delta.
\label{eq:app_McDiarmid_bound}
\end{equation}
Observe that the bound does not depend on $N_S$ --- the optimal strategy is to generate only one-shot class vector for each $U$, maximising $N_U \mapsto N / 3$ and obtaining
\begin{equation}
    N \ge \frac 3{\epsilon^2} \left( 1 + \sqrt{\ln(1/\delta)}\right)^2.
\end{equation} 
It guarantees an $\ell_2$ error at most $\epsilon$ with confidence at least $1-\delta$.  Crucially, this bound improves the one in Eq. \eqref{eq:app_B.H.C._Ntot} because it removes the dependence on the cardinality $|\{\mathcal C _\mu\}_{\mu}|$ of the set of equivalence classes, whereas its dependence is only linear in the order of the correlator, i.e. 3.
\end{document}